\documentclass[epj]{svjour}
\usepackage{graphics}
\usepackage{psfrag}
\usepackage{epsfig}
\begin{document}
\title{\hfill{\small FZJ--IKP(TH)--2006--26} \\
Towards an understanding of the light scalar mesons}
\author{C. Hanhart
}                     
%
%
\institute{Institut f\"ur Kernphysik, Forschungszentrum
J\"ulich, D-52425 J\"ulich, Germany}
\date{Received: date / Revised version: date}
%
\abstract{Although studied for many years the nature of the light
scalar mesons remains controversial. Here we shall present
a method, applicable for s-wave states located close to a
threshold, that allows one to quantify the molecular part of a
given state. When applied to the $f_0(980)$ a dominance of the molecular
component is found. In the second part we show that requirements of
field theoretic consistency and chiral symmetry, when applied to the
scattering of light pseudo--scalars, naturally lead to the appearance of
dynamical poles in the scalar sector. A program is proposed on how to
further investigate experimentally the mixing between these dynamical
states and possible genuine quark states.
} 
\maketitle
\section{Introduction}

At first glance it comes as a surprise that
the lowest scalar excitations of QCD are still not 
fully understood. First, there
has been a long debate on how many scalar states there
are below 1 GeV. Besides the well established isovector
$a_0(980)$ and the isoscalar $f_0(980)$ there was experimental
as well as theoretical evidence
for two more states, namely the $f_0(600)$ --- often
called the $\sigma$ meson --- and the isodoublet $\kappa$.
Together these states could fill the lowest scalar
nonet.

 Although recent efforts in
dispersion theory in combination with chiral
perturbation theory unambiguously determined
the existence as well as the position 
of the $\sigma$--pole \cite{gilberto}, 
the discussion of the very nature of this state
and its relatives is far from settled.
Analyses can be found in the literature  that identify
these structures with
 conventional $q\overline{q}$ states (see e.g.\ Refs.\ 
\cite{Morgan}) --- in some analyses with a sizable
admixture from the continuum \cite{vanb,toernqvist} --- compact $qq-\bar{q}\bar{q}$ states
\cite{Achasov,schechter} or loosely bound $\bar K K$ molecules
\cite{Weinstein,oop}.

What is  clearly called for is a study 
that identifies those cases when the nature
of a particular state can be read off from an
experimental observable. Based on an old proposal
by Weinberg \cite{wein},
a first step in this direction was taken  in Ref. \cite{evidence} --- specifically,
the original argument was extended to also
allow for the presence of inelasticities.
The conditions where this method can be applied 
were found to be
\begin{itemize}
\item the state must be an $s$--wave 
with respect to the continuum states\footnote{Not to be obscured
with the quantum numbers with respect to the quark
constituents.};
\item the binding energy $\epsilon$ must be {\it much}
smaller than any intrinsic scale of the problem; 
\item any inelastic threshold must be 'far away' (in units
of the binding energy) from the elastic threshold of interest.
\end{itemize}
It was the central finding of Ref. \cite{evidence}
that also for inelastic interactions the value of the
effective coupling of a resonance to the continuum
state of interest\footnote{To be more specific: what is meant
is the corresponding residuum at the resonance pole.} is a direct measure of the molecular
component; especially, its value gets maximum (up to higher order
corrections) in the case
of a pure molecule. 

For the derivation of the above mentioned result we refer to
Ref. \cite{evidence}. In this brief note we will focus more on a
discussion of why this works and how to further exploit this insight.
Thus, in the next section, the role of the effective coupling will be
discussed and the scheme will be applied to the $f_0(980)$.  In
section 3 we argue that a prominent molecular structure of the
light scalar mesons emerges quite naturally from the properties of the
meson--meson scattering amplitude near threshold controlled by chiral
symmetry. In section 4 we shall briefly comment on possible further
experiments to investigate the mixing of the light scalar mesons with
the (heavier) quark states.

\section{How does this work?}
\label{hdtw}


For simplicity let us focus on a situation where two spin zero mesons
 of mass $m$ couple to a single, isolated resonance state.
The possible presence of inelasticity will be discussed later.
The two point function $g(s)$ for the resonance state
may then be written as
\begin{eqnarray} 
\nonumber
g(s) &=& \frac1{s-M_0^2-i\Sigma(s)}+\mbox{regular terms} \ \\
&=& \  \frac{Z}{s-M^2}+\mbox{regular terms}
\end{eqnarray} 
where  $M_0$ ($M^2$), $\Sigma$, $Z$, denote the bare (physical) mass,
 the self energy, and the wave function renormalization of the state of interest.
It is straightforward to show that (for a non--relativistic system) the
$Z$ factor measures the bare state admixture of the physical state \cite{wein,evidence}.
Thus, $Z=0$ ($Z=1$) corresponds to a pure molecule (compact state).

 By assumption the binding energy
is much smaller than any intrinsic scale of the problem --- 
assumed especially to be smaller that the inverse of the range of forces. In this situation
the vertex that couples the state of interest to the continuum can
be safely assumed to be point--like 
and the self energy is just the standard scalar loop function
times a strength parameter that we will call $G$.
One then easily derives
\begin{equation}
\frac{1}{Z}-1 = \frac{G}4\sqrt{\frac{m}{\epsilon}} \ + \ \mathcal{O}\left(\epsilon R\right) \ ,
\label{main}
\end{equation}
where $R$ denotes the range of forces.
 Weinberg applied
this to the deuteron, where $R\sim 1/m_\pi$. For the scalar
mesons, on the contrary, we look at the scattering of two pseudo--scalar
mesons. Thus the lightest particle that can be exchanged in the $t$--channel
is the $\rho$ meson from which we get $R\sim 0.25$ fm, which is of the order
of the extension of conventional mesons.
 Under the conditions assumed the $1/\sqrt{\epsilon}$ term should
dominate the right hand side of Eq. (\ref{main}) thus allowing one to express the
effective coupling $G$ in terms of $Z$, which 'measures' the nature
of the state.  On the other hand the effective coupling $ZG$ is --- in
principle --- a measurable quantity: it is the residue of the
scattering matrix at the resonance pole and it can be related to
 the
scattering length and the effective range for the meson--meson
scattering. This follows directly from matching the expression
for the meson--meson scattering matrix---$Gg(s)$---to the corresponding effective range expansion 
\begin{eqnarray}
Gg(s)&=&\frac{G}{s-M^2+iMG\sqrt{s-4m^2}/2} \nonumber \\
    &=& \frac1{2m}\left(\frac{G/2}{\epsilon+k^2/m+G/2(ik+\sqrt{m\epsilon})}\right)
+{\cal O}\left(\frac{k^2}{m^2}\right) \nonumber \\
&=& -\frac1{2m}\left(\frac{1}{1/a+r/2\, k^2-ik}\right)+{\cal O}\left(\frac{k^2}{m^2}\right) \ ,
\label{gel}
\end{eqnarray}
which leads (up to higher orders) to
\begin{eqnarray}
-\frac{1}{a}=\frac{2\epsilon}{G}+\sqrt{m\epsilon} \ , \quad -\frac12\, r = \frac{2}{Gm} \ .
\end{eqnarray}
Using Eq. (\ref{main}) one may equivalently express $a$ and $r$ directly in terms of $Z$.

\begin{figure}
\psfrag{@}{{\huge $\sqrt{s}$}}
\resizebox{0.45\textwidth}{!}{%
  \includegraphics{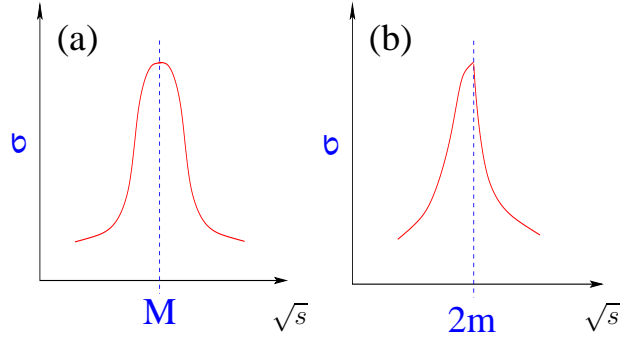}
}
\caption{Typical resonance signals for a 
genuine state (a) and a dominant molecular state (b).}
\label{fig:1}       
\end{figure}

Note that the coupling $G$ controls the relative importance
of the term non--analytic in $s$ to that analytic in $s$.
Since the only source of a non--analyticity are the unitarity cuts,
this is exactly what controls the amount of molecular admixture.
For a recent discussion of $s$-wave thresholds on amplitudes
we refer to Ref. \cite{rosner} and references therein.
We shall come back to this below.

For later use it is convenient to introduce $g_{eff}$, defined as
\begin{eqnarray}
\nonumber
\frac{g^2_{eff}}{4\pi}&=& Z8m^2G \\ \nonumber &=& 32(1-Z)m\sqrt{\epsilon m}\le 32m\sqrt{\epsilon m} \ .
\label{geff}
\end{eqnarray}
 It is this effective coupling that controls the resonance coupling in
inelastic reactions like $\phi \to \gamma \pi\pi$.  This reaction will
be discussed in more detail below. It is important to observe that the
formalism sketched here gives a results that is intuitively clear: the
larger the effective coupling of the physical resonance to the
continuum the larger the probability to find the continuum state in
the physical state or, stated differently, the larger its molecular component.
Eq. (\ref{geff}), however, makes an even stronger statement: the maximum
coupling is constrained from above and the value of this maximal
coupling is controlled solely by the binding energy. Below we
will call the model with $g^2_{eff}/4\pi=32m\sqrt{\epsilon m}$
the {\it naive molecular model}.

What changes now if we introduce an inelastic channel? By assumption this
new channel is not allowed to introduce any new small scale into
the problem. Thus we may assume that the inelastic threshold
is, when measured in units of the binding energy, very far away.
Then its leading effect is to introduce a constant (or at most weakly
energy dependent) imaginary
part $i\Gamma$ to the denominator of $g(s)$, or --- equivalently --- to the
self energy. 
\begin{equation}
g(s)=\frac1{s-M^2+iMG\sqrt{s-4m^2}/2+i\Gamma}  \ .
\label{ginel}
\end{equation}
Thus, the appearance
of $\Gamma$ does not change the relative importance of the $s-M^2$
piece and the $iMg\sqrt{s-4m^2}/2$ piece, and consequently
$G$ still measures the amount of molecular admixture with respect
to the elastic channel \cite{evidence}. A discussion on the
subleading corrections is provided in Ref. \cite{kerb}.

Eq. (\ref{ginel}) is nothing but the standard Flatt{\'e} form
used for the parameterization of resonance signals near 
thresholds \cite{Flatte}. As stated above, for a state with
a negligible continuum admixture, the term linear in $G$
of Eq. (\ref{ginel}) can be neglected and the resulting
distribution for the resonance is that of a standard Breit--Wigner.
This is sketched in Fig. \ref{fig:1}a. On the other hand, if the
state is predominantly a molecule, it is the $G$--term that
controls the dynamics near the meson--meson threshold and
the variation of $s$ in the first term may be neglected.
The resonance signal then produces a very pronounced cusp structure, for which a typical
case is shown in Fig. \ref{fig:1}b. (Many different shapes are possible; 
see the discussion in Ref. \cite{Flatte}.)
It therefore appears a straight forward task
to use measured Flatt{\'e} distributions to extract
$G$ and deduce from this the molecular admixture. Unfortunately,
due to a scale invariance of the amplitude, an extraction
of the absolute value of $G$ directly from the mass distributions
 is very difficult. (See Ref. \cite{flattecrit} for a detailed
discussion.) As we shall see, inelastic processes might be
more useful here.

Thus, qualitatively one comes to the conclusion that the
more distorted a Breit--Wigner distribution gets through
the opening of a production threshold the more molecular
component there is in the corresponding resonance. In
Ref. \cite{evidence} we tried to put this on more quantitative
grounds. 
Thus the strongly asymmetric mass distribution seen for
the $f_0(980)$ by the BES collaboration \cite{bes} in the reaction
$J/\Psi\to \phi \pi\pi$ already provides strong evidence
for a predominantly molecular composition of the $f_0$.

\section{The physics of light scalar mesons}

It is surprising that the lowest scalar
excitations in QCD seem to be of quite complicated structure --- even
more complicated than the higher excited states, where we assume to
find the $\bar qq$ states. In this section we will argue that 
the appearance of low--lying dynamical states is a natural consequence
of chiral symmetry. Note, here 'natural' should not be confused
with necessary. In this context see discussion in Ref. \cite{josehiggs}.

The observation essential for our argument is the energy dependence 
of the $\pi\pi$ scattering amplitude near its threshold. It
reads in the scalar--isoscalar channel \cite{wein2}
\begin{equation}
V_{\pi\pi} = (s-m_\pi^2/2)/f_\pi^2 \ .
\label{vpipi}
\end{equation}
Corrections to this expression can be calculated using 
chiral perturbation theory in a controlled way \cite{gasser}.

One may ask if such an energy dependence near threshold
can emerge solely from an $s$--channel resonance. The answer
is ``no'' for two reasons and may be read off the corresponding
scattering matrix directly. The scattering potential
for scattering through a resonance has the form
\begin{equation}
V_R = G_R^2/(s-m^2) \ ,
\label{vr}
\end{equation}
thus, matching to $V_{\pi\pi}$ would
give first of all a relation between the masses scalar resonances 
and $f_\pi$ that looks very counter intuitive; but, even
more importantly, for $V_R$ to employ the energy dependence
of $V_{\pi\pi}$ calls for 
\begin{equation}
G_R \propto \sqrt{s-m_\pi^2/2} \ .
\label{gr}
\end{equation}
In order to give the correct energy dependence near 
threshold the amplitude must contain an unphysical
branch point. Clearly, in elastic $\pi\pi$ scattering
this will not show up, because only $G_R^2$ appears.
Production amplitudes, however, are linear in $G_R$ and thus
one would start to notice the unphysical cut in calculations
based on Eq. (\ref{vr}). 
Although not allowed by analyticity, several works
use the vertex function of Eq. (\ref{gr}) (see e.g. Ref. \cite{toern})
or similar ones  (see e.g. Ref. \cite{vanbev}).

To cure this problem one option would be to use
the linear sigma model. There a four pion
contact term is present in addition to the $\sigma$ pole 
diagrams and a threshold amplitude of the form of \ref{vpipi}
emerges naturally. Then one finds that the contact term 
as well as the $t$-- and $u$-- channel exchanges play a 
prominent role in the dynamics below 1 GeV~\cite{achasovalt}.
An alternative and theoretically more appealing approach (see
discussion in Ref. \cite{ulfalt}) is chiral perturabation
theory. There only pions (and in the SU(3) extension also 
kaons and eta mesons) appear as dynamical fields that 
interact through contact interactions constructed
consistent with chiral symmetry. The corresponding
leading order Lagrangian automatically
gives a potential of the form of Eq. (\ref{vpipi}) --- see
Ref. \cite{gasser} and references therein.  
In this scheme  there is a strongly energy
dependent, non--resonant $\pi$-$\pi$ interaction to be included in the
theory, whose $s$--dependence is so strong that the tree--level
potential of Eq. (\ref{vpipi}) hits the unitarity bound already at
quite low energies. Or, stated differently, a properly unitarized
amplitude will naturally employ a pole in the complex plane to prevent
the amplitude from growing beyond what is allowed by unitarity. This
pole should be identified with the $f_0(600)$ --- the $\sigma$ meson.
This picture was introduced in Refs. \cite{ulfalt,schechter} and was further
supported by studies within unitarized chiral perturbation theory
\cite{pd}. For a recent review see Ref. \cite{joserev}.

So far the argument was given for $\pi\pi$ scattering
only, however studies within both unitarized
chiral perturbation theory discussed above as well
as phenomenological models \cite{janssen} show that
the same holds for the scattering of all the pseudo--Goldstone 
bosons with each other. As a consequence,
in addition to the lowest pole in the $\pi\pi$ channel,
coupling to the $\bar KK$ system leads to a pole close to the $\bar KK$ threshold,
identified with the $f_0(980)$.
Also in the $\pi \eta$-$\bar KK$
coupled system a pole appears close to the $\bar KK$ threshold,
interpreted as $a_0(980)$ and in the $\pi K$ channel the
$\kappa$ pole appears.

To summarize this part, we find as a natural consequence
of the analytic properties of the scattering
amplitude for the scattering of the ground state
pseudo--scalar mesons with each other that dynamical
poles with scalar quantum numbers are produced.
This insight is fully in line with the experimental
evidence that the $f_0(980)$ is predominantly of
dynamical origin. What remains to be seen is 
the mixing pattern of those molecular states with
the non--molecular states. In the rest of this
presentation we shall argue that exploiting
the matrix element for scalars coupling to
a photon and a vector meson in various kinematic
regimes is what should provide important information
in this direction.

\begin{figure*}[t]
\begin{center}
\begin{tabular}{cccc}
\epsfig{file=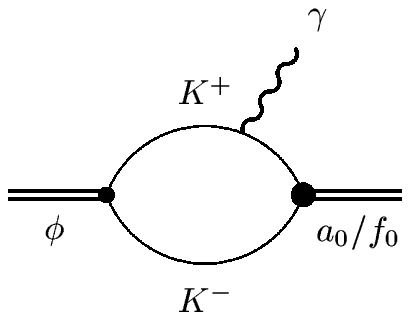,width=3.5cm}&
\raisebox{-6mm}{$\epsfig{file=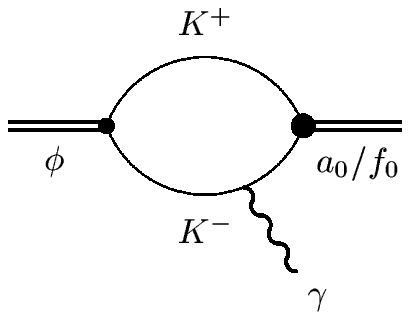,width=3.5cm}$}&
\epsfig{file=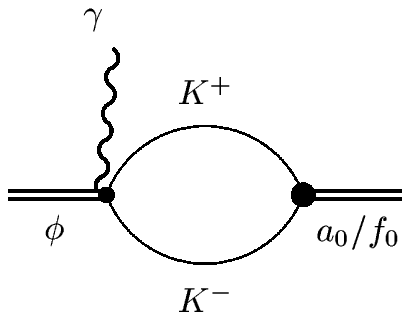,width=3.5cm}&
\epsfig{file=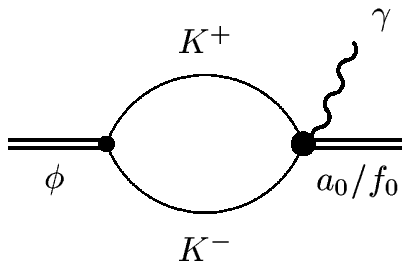,width=3.5cm}\\
(a)&(b)&(c)&(d)
\end{tabular}
\end{center}
\caption{Kaon loop contributions  to the radiative decay amplitude.}
\label{loopdiags}
\end{figure*}

\section{$\phi\to \gamma \pi\pi$ and possible further experiments}

In recent years a lot of experimental as well as theoretical effort
went  into studies of
the reactions $\phi\to \gamma \pi\pi$ and $\phi\to \gamma\pi\eta$,
both believed to shed light on the nature of the light scalar 
mesons $f_0$ and $a_0$, respectively.
Following the reasoning of Ref. \cite{sraddec} and that of the previous
sections, in this section
we shall argue that in these reactions
the molecular component of the scalar mesons was measured. In addition we shall
show that looking
at the same matrix elements for different masses of the vector
meson will allow one to study the mixing of this molecular component
with possible compact states.

There is strong experimental evidence that 
the reaction $\phi \to \gamma \pi\pi$ in the
upper part of the available phase space runs predominantly
through a kaon loop followed by a resonance formation into the
$f_0(980)$. The kaon loop dominance was observed by
Achasov and Kiselev (see Ref. \cite{achasovlatest} and references therein). The
reason why one can in the reaction $\phi\to \gamma \pi\pi$ almost  'see' the kaon 
loops in the spectrum is the following: 
gauge invariance demands the transition amplitude for $\phi\to \gamma \pi\pi$
to vanish for vanishing values of $\omega$ --- the energy of the outgoing photon.
As a consequence the $\phi$ decay rate has to scale as $\omega^3$ at the upper end
of phase space. On the other hand, 
the experimental spectrum of Refs. \cite{CMD,KLOE} is (almost) identical to
that expected from an undistorted $f_0$ spectral function. What
looks contradictory at first glance is quite natural, since the $\bar KK$
threshold is very close to the mass of both the $\phi$ and the $f_0$.
Consequently there is a pronounced cusp structure in the kaon loop
that effectively compensates the mentioned $\omega^3$ suppression
at the upper end of phase space.

In the previous section we argued that the effective coupling 
of the scalar mesons to kaons is  a measure for the importance
of the molecular component of the $f_0(980)$. We now see that
the kaon loop dominates the transition rate $\phi\to \gamma \pi\pi$.
We therefore have to conclude that the radiative decay of
the $\phi$ into a pion pair measures the molecular component 
of the scalar meson. Even more, we may use the naive molecular
model introduced above to estimate the rate for $\phi\to \gamma\pi\pi$
based on Eq. (\ref{geff}) with $Z=0$ and a typical value of $\epsilon=10$ MeV.
Then we get \cite{sraddec} $\Gamma \sim 0.6$ keV to be compared to the
experimental value of $0.4$ keV \cite{KLOE}. This we interpret
as strong evidence in favor of a prominent molecular
structure of the $f_0$. This conclusion is in line
with the results of calculations within the unitarized 
chiral perturbation theory for the $\phi$ radiative decay \cite{Oset,Markushin,Oller}.

\begin{figure*}[t]
\epsfig{file=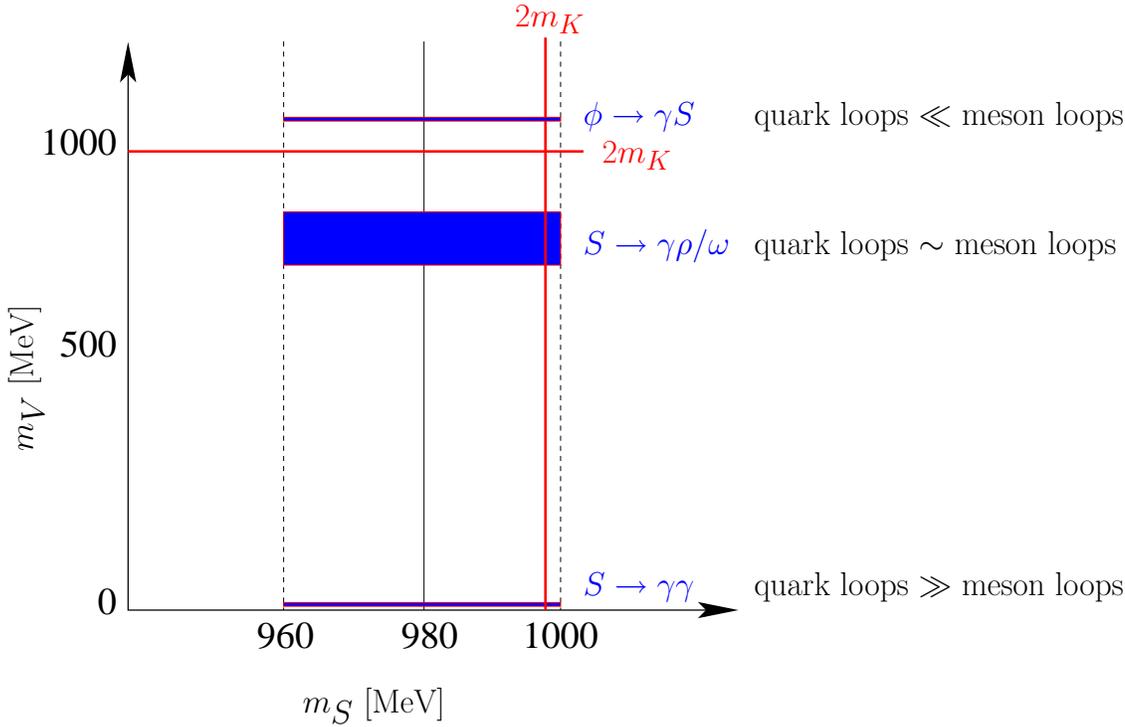, width=15cm}
\caption{\it Illustration of various kinematical regimes probed by the
decays involving scalars.}
\label{melo2}
\end{figure*}

It should be mentioned that the interpretation of the data for
$\phi\to \gamma \pi\pi$ is still controversial and the above picture
is not yet fully accepted.  In Ref.~\cite{CIK} it is claimed that
there should be a strong suppression of the $\phi \to \gamma f_0/a_0$
branching ratio for the scalars in case they are loosely bound
molecules as compared to point--like scalars that correspond to compact
quark states, ($10^{-5}$ {\it vs} $10^{-4}$).  A study by Achasov et
al. \cite{AGS}, where the finite width of scalars was taken into
account, arrived at the same conclusion.  Thus, the authors of
\cite{CIK} and \cite{AGS} stress that data for this branching ratio
should allow to prove or rule out the molecular model of the
scalars. However, this conclusion is based on a confusion between the
notion of a wave function and that of a vertex function. Indeed, the
relevant scale in the spatial extension of a wave function of a
molecule is indeed the binding energy: the smaller the binding the
larger the extension. What enters, however, in the loops depicted in
Fig. \ref{loopdiags} is not the wave function but the vertex
function, whose scale is set by the range of forces --- for a molecule
made of two pseud scalars this is of the order of the mass inverse of
the $\rho$ meson, much smaller than what derives from the binding
energy. The wave function, on the other hand, is proportional to the
vertex function times the two--meson propagator (see Eq. (22) in
Ref. \cite{phidecmol}). This very propagator, however, is already
included explicitly in the integral for the  kaon loop\footnote{For
more details on this discussion see the comment on Ref. \cite{sraddec}
in Ref. \cite{acom} and our reply to this in Ref. \cite{acomcom}.}.

Recently an attempt was made to set up an analysis
for the data on $\phi\to \gamma \pi\pi$ that avoids
the use of the kaon loops \cite{modelindep}.
The authors parameterize the $\phi\to\gamma \pi\pi$ 
amplitude as a polynomial in $s$ and succeeded
in fitting the $\phi$--decay spectra once a sufficient
number of terms was included in the polynomial. 
This one might interpret as an indication that
the kaon loops are not of as high relevance as 
indicated in the previous paragraphs. However, 
a polynomial in $s$ will not have the non--analytic
pieces that are provided by the kaon loops.
A truly model--independent analysis would therefore
include both the mentioned polynomial as well as
the kaon loop. A fit then needs to decide how 
much loop is really necessary. Given the comments
above it is likely that such a fit will call for
a quite small contribution from the terms analytic in $s$.

Thus far we have argued that the $\phi$ radiative
decay measures the molecular component of 
the scalar mesons. What remaines unanswered so far
is the amount of mixing of molecular states with 
genuine quark states. As argued in Ref. \cite{sraddec},
further insight into this issue can be gained from a systematic
study of decays of scalar mesons into a photon and a vector meson.
The physics behind this is quite easy: the reason why the $\phi$ radiative
decays have large kaon loop contributions is the proximity
of the $\bar KK$ threshold to both the mass of the $\phi$ as well
as of the scalars of interest here. Looking at the radiative
decays of the scalars allows us to study the same matrix element 
in different kinematics: we can now change the mass of the vector
meson away from that of the $\phi$ to that of the $\omega$ and $\rho$, or 
even to $m=0$ for the decay $s\to \gamma \gamma$. As explained
above, the unitarity cut related to the $\bar KK$ threshold 
introduces a significant energy dependence to the kaon loop.
It gets smaller the further we move away from the threshold. 
On the other hand, quark loops do not feel the kaon threshold and
therefore a much weaker energy dependence is expected for these. 
This means, that as we move from the the $\phi s \gamma$ vertex to 
the $(\rho\omega) s \gamma$, and eventually the $\gamma s \gamma$ vertex, 
it should be straightforward to disentangle the kaon loop part
of the matrix element from the quark lines. 
This logic is sketched in Fig. \ref{melo2}\footnote{Whenever
quark loops and meson loops are considered simultaneously there
is the possibility of double counting. However, here this is 
not the case since the relavant meson loops are finite.}.

In oder to be more quantitative, particular models need to be employed
for the quark loop contributions. For details we refer to
Ref. \cite{sraddec}. It is important to stress that already the
naive molecular model, as sketched in section \ref{hdtw}, gives
values for the decays $s\to \gamma\gamma$ that are of the right
order of magnitude.

\section{Summary}

To summarize, we argued that the available data --- especially
that on the $\phi$ radiative decays --- are compatible with a model
that assumes the $f_0$ to be predominantly of molecular nature.
In order to determine the mixing scheme of this molecule and
its $SU(3)$ relatives with possible quark states a series of
experiments that studies the radiative decays of the scalars
was proposed.

\vspace{0.5cm}

\noindent
{\bf Acknowledgment}

\noindent
The results presented were found in collaboration with M. Evers,
J. Haidenbauer, Yu. S. Kalashnikova, S. Krewald, A. Kudryavtsev,
A. V. Nefediev --- thanks to all for this very fruitful and enlightening 
collaboration.
The author also thanks J. Durso for numerous editorial remarks
and U.--G. Mei\ss ner for critical reading.
 This research was supported in part
by the grant DFG-436 RUS 113/733.

I would like to thank the organizers of the QNP conference,
especially Felipe J. Llanes Estrada, for a informative, educating and 
entertaining conference.

\end{document}